\documentstyle[preprint,aps,epsfig]{revtex}
\tighten
\draft
\begin{document}

\bibliographystyle{prsty}
\preprint{RUB-TPII-05/97}
\title{Flavor structure of the octet magnetic moments}

\author{Hyun-Chul Kim
\footnote{permanent address: Department of Physics, 
Pusan National University, Pusan 609-735, Korea},
Michal Prasza\l owicz
\footnote{On leave of absence from Institute of Physics,
Jagellonian University, Cracow, Poland},
Maxim V. Polyakov
\footnote{On leave of absence from PNPI, Gatchina,
St.Petersburg 188350, Russia},
and
Klaus Goeke
}
\address{
Institut f\"ur Theoretische  Physik  II, \\  Postfach 102148,
Ruhr-Universit\" at Bochum, \\
 D--44780 Bochum, Germany  \\  }
\date{June, 1998}
\maketitle
\begin{abstract}
We use the chiral quark-soliton model to identify all symmetry breaking
terms linear in $m_{\rm s}$ and investigate the strange magnetic moment
in a ``model-independent'' way.  Assuming hedgehog symmetry and employing
the collective quantization, we obtain the most general expression for the
flavor-singlet and flavor-octet magnetic moments in terms of seven
independent parameters. Having fitted these parameters to the experimental
magnetic moments of the octet baryons, we show that the strange magnetic
moment turns out to be positive. The best fit obtained by minimizing
$\chi^2$   assuming 15\% theoretical accuracy yields:
$\mu^{({\rm s})}_{\rm N} = (0.41 \pm 0.18 )\; \mu_{\rm N}   $.
\end{abstract}

\newpage

The strangeness content of the nucleon has been a hot issue.  The
analysis of the $\pi N$ sigma term $\Sigma_{\pi N}$ implies a
non negligible scalar density of the strange quark in the
nucleon~\cite{GasserLeutwylerSainio}, once known as ``{\em $\Sigma_{\pi
N}$ puzzle}''.  In the axial channel, deep inelastic muon scattering
conducted by European Muon Collaboration (EMC) almost ten years ago
indicates that there is a sizeable contribution of the strange quark
spin $\Delta s$ to the proton spin, which was once called ``{\em spin
crisis}''.
The measurement of elastic $\nu p/\bar{\nu} p$ cross section  at
Brookhaven (BNL experiment 734), from which the strange axial form
factor was extracted, came to the more or less same conclusion.  In the
vector channel the strange magnetic moment presents another
key of understanding the strangeness content of the nucleon~\cite{km}.
Great deal of theoretical effort has been put into the investigation of
the strange magnetic moment and strange radius
\cite{Ja}\nocite{Pa,pw,Ko,Co,Ho,mu,fo,Le,Ha,Mus,Ki,Me}--\cite{HongParkMin},
which gave the range of possible values of the strange magnetic moment:
$(-0.75\rightarrow 2.2)\:\mu_{\rm N} $.  This indicates that the strange
properties of the nucleon in the vector channel are still poorly understood.

Very recently, the SAMPLE collaboration has announced the first
measurement of the strange magnetic form factor $G^{s}_{M}$ at
$Q^2=0.1\;\mbox{GeV}^2$~\cite{SAMPLE}:
\begin{equation}
G^{({\rm s})}_{\rm M}\;=(\;+0.23\pm 0.37 \pm 0.15 \pm 0.19)\, \mu_{\rm N}.
\end{equation}
Although the central value for $G^{({\rm s})}_{\rm M}$ is positive, it is
still too early to draw any firm conclusion, because of large experimental
errors. One has to wait for other experiments which are presently either
being conducted or planned at various electron accelerators
\cite{CEBAF1,CEBAF2,CEBAF3,Mainz,MIT1,MIT2}. It is therefore of importance
to make theoretical predictions for this quantity.

Recently \cite{KPBG,WaKaya,KimPraGo} magnetic moments of baryons have been
studied within the framework of the chiral quark-soliton model ($\chi$QSM)
-- see review article \cite{review} and references therein.
In Ref.~\cite{KimPraGo} the "model-independent" way of analyzing
magnetic moments has been proposed, in which the intrinsic dynamical
parameters of the chiral quark-soliton model were fitted to the
experimental data of the octet magnetic moments.  In this way the
magnetic moments of the baryon decuplet have been predicted.  Actually,
the results for the $\mu_{\Omega^-}$ and $\mu_{\Delta^{++}}$ are in a
remarkable agreement with the data.  Encouraged by these results, in
the present work we apply the same procedure to predict the flavor
magnetic moments. This requires the evaluation of the singlet part of
the magnetic moment operator, which has not been calculated in
Ref.~\cite{KimPraGo}.

Our present investigation is in line with the recent work by Hong,
Park, and Min~\cite{HongParkMin}, which studied the strange magnetic
moment of the nucleon within the framework of the chiral bag model
using the "model-independent" method analogous to that of
Ref.~\cite{KimPraGo}.  However, in Ref.~\cite{HongParkMin} the
singlet magnetic moment does not contain the SU(3) symmetry breaking
contribution; also the triplet and octet parts of the magnetic moments
do not contain the symmetry breaking pieces which are present in our
approach.  Hence, we aim in this letter at investigating the flavor
structure of the magnetic moments of the SU(3) baryons, taking
consistently into account the  contribution of the mass of the strange
quark up to the linear order. We also discuss flavor structure of
the magnetic moments for other members of the baryon octet.

Although it turns out that our numerical results for neutron and
proton  are very close to the ones of Ref.~\cite{HongParkMin}, we
discuss in more detail the reliability of the whole approach by
calculating the flavor magnetic moments for all baryons in the octet.
We are confident that $\mu^{(\rm s)}_{\rm N}$ is positive and of the
order 0.2$-$0.6~$\mu_{\rm N}$.

The vector form factors of the SU(3) baryons are defined by
the matrix elements of the $\mbox{U}_V(3)$ vector currents:
\begin{equation}
\langle B(p') | V^{(a)}_\mu | B(p)\rangle \;=\;
\bar{u}_B(p') \left[\gamma_\mu F^{(a)}_1 (q^2) + i\sigma_{\mu\nu}
\frac{q^\nu}{2M_N} F^{(a)}_2 (q^2)\right] u_B (p)
\label{Eq:def}
\end{equation}
with $a=0,3,8$.
The $\mbox{U}_V (3)$ currents are defined as follows:
\begin{equation}
V^{(0)}_\mu\;=\; \frac{1}{3} \bar{\psi} \gamma_\mu \psi, \;\;\;
V^{(3)}_\mu\;=\;  \overline{\psi} \gamma_\mu \lambda^3 \psi,\;\;\;
V^{(8)}_\mu \;=\; \overline{\psi} \gamma_\mu \lambda^8 \psi .
\end{equation}
$q^2$ in Eq.(\ref{Eq:def}) denotes the square of the four momentum
transfer.The magnetic Sachs form-factor  $G^{(a)}_{\rm M}$ can be
defined in terms
the vector form-factors $F^{(a)}_1$ and $F^{(a)}_2$ in the following way:
\begin{eqnarray}
G^{(a)}_{\rm M} &=& F^{(a)}_1(q^2) + F^{(a)}_2 (q^2) .
\end{eqnarray}
In the non relativistic limit
$G^{(a)}_{\rm M}$ can be related to the time and space components
of the $\mbox{U}_V(3)$ vector currents, namely:
\begin{eqnarray}
\langle B(p') | V^{(a)}_i | B(p)\rangle &=& \frac{1}{2M_N}
G^{(a)}_M (q^2) i\epsilon_{ijk} q^j \langle s' | \sigma_k |s\rangle,
\label{Eq:form}
\end{eqnarray}
where $\sigma_k$ denotes Pauli spin matrices while $|s\rangle$ is the
corresponding spin state of the baryon.
The magnetic moments $\mu^{(a)}$ corresponding to the vector currents
are identified with $G^{(a)}_{\rm M} (0)$.

The matrix elements given above are related to the correlator
\begin{equation}
\label{corf}\langle 0|J_{B}({\bf x},T)\bar \psi \gamma _\mu \hat Q\psi
J_{B}^{\dagger }({\bf y},0)|0\rangle
\end{equation}
at large Euclidean time $T$ . The baryon current $J_B$ can be constructed
from $N_{{\rm c}}$ quark fields,
\begin{equation}
J_B=\frac 1{N_{{\rm c}}!}\varepsilon ^{i_1\ldots i_{N_{{\rm c}}}}\Gamma
_{SS_3II_3Y}^{\alpha _1\ldots \alpha _{N_{{\rm c}}}}\psi _{\alpha
_1i_1}\ldots \psi _{\alpha _{N_{{\rm c}}}i_{N_{{\rm c}}}}
\end{equation}
$\alpha _1\ldots \alpha _{N_{{\rm c}}}$ are spin--isospin indices, $%
i_1\ldots i_{N_{{\rm c}}}$ are color indices, and the matrices $\Gamma
_{SS_3II_3Y}^{\alpha _1\ldots \alpha _{N_{{\rm c}}}}$ are taken to endow the
corresponding current with the quantum numbers $SS_3II_3Y$. $%
J_B(J_B^{\dagger })$ annihilates (creates) the baryon state at given time $T$.
Taking into account the rotational corrections to the order ${\cal
O}(1/N_{{\rm c}})$ and ${\cal O}(m_{{\rm s}})$ (for details, see for
example Ref.~\cite{review}), the general expressions for the collective
magnetic moment operators $\hat{\mu}^{(a)}$ can be written as follows:
\begin{eqnarray}\label{Eq:mu}
\hat{\mu}^{(0)} &=& \frac13 w_3 \hat{S}_3 + \frac{\sqrt{3}}{3}
m_{\rm s}\, \left(w_5 - w_6\right) D^{(8)}_{83}, \nonumber \\
\hat{\mu}^{(a)} &=&\left( w_1^1+m_{{\rm s}}w_1^2\right)
\;D_{a3}^{(8)}\;+\;w_2d_{pq3}D_{ap}^{(8)}\cdot \hat S_q\;+
\;\frac{w_3}{\sqrt{3}}D_{a8}^{(8)} \hat S_3   \\
&+&m_{\rm s}
\left[ \frac{w_4}{\sqrt{3}}d_{pq3}D_{ap}^{(8)}D_{8q}^{(8)}+w_5
\left( D_{a3}^{(8)}D_{88}^{(8)}+D_{a8}^{(8)}D_{83}^{(8)}\right)
\;+\;w_6\left( D_{a3}^{(8)}D_{88}^{(8)}-D_{a8}^{(8)}D_{83}^{(8)}\right)
\right] \nonumber
\end{eqnarray}
with $a=3$ or 8. $\hat S_a$ stands for the operator of the generalized
spin acting on the angular variable $R(t)$~\cite{Blotzetal}.
$D_{ab}^{({\cal R)}}(R)$ denotes the SU(3) Wigner matrix in the
representation ${\cal R}$ and  $m_{\rm s}$ is the mass of the strange
current quark.   The parameters $w_i$ depend on the dynamics of a specific
hedgehog model.

Eq.(\ref{Eq:mu}) is different from Eq.(5) of Ref.~\cite{HongParkMin}
by terms proportional to $m_{\rm s}(w_5-w_6)$.  In
Ref.~\cite{HongParkMin} $w_5=w_6$, and as a result the singlet magnetic
moment operator $\hat\mu^{(0)}$ is degenerated in the octet and
decuplet.  Also $\hat\mu^{(3)}$ and $\hat\mu^{(8)}$ get extra
contributions in the present case, where $w_5 \ne w_6$.  Numerically  these
terms are not very important in the case of the octet magnetic moments.
This is, however, visible only {\em a posteriori}.  Magnetic moments
are  particularly sensitive to the SU(3) symmetry breaking, therefore
careful analysis of all possible symmetry breaking terms is of quite
importance.

The operator (\ref{Eq:mu}) has to be sandwiched between the octet
 collective wave functions. However, strictly speaking, octet
baryon wave functions are no longer pure octet states:
\begin{equation}
\Psi_B (R) \;=\; \Psi^{(8)}_B (R) + m_{\rm s}\,c_{\overline{%
10}}^B\ \Psi^{(\overline{10})}_B (R)
+m_{\rm s}\,c_{27}^B\ \Psi^{(27)}_B (R),
\label{state8}
\end{equation}
where
\begin{equation}
c^{B}_{\overline{10}} \;=\; c_{\overline{10}} \, \sqrt{5} \
\left[ \begin{array}{c}  1 \\ 0 \\ 1 \\ 0
\end{array} \right],\;\;\;\;
c^{B}_{27} \;=\; c_{27} \
\left[ \begin{array}{c} \sqrt{6} \\ 3 \\ 2 \\  \sqrt{6}
\end{array}\right]
\label{Eq:g2}
\end{equation}
in the basis of $[N,\ \Lambda,\ \Sigma,\ \Xi]$.
The coefficients $c_{\overline{10}}$ and $c_{27}$ are defined in
Ref.~\cite{KimPraGo} and depend on the inertia parameters $I_i$ and
$K_i$ $(i=1,2)$ calculable in the model. For the purpose of the present
"model independent" analysis we shall use the following approximate
equality:
\begin{equation}\label{ratios}
\frac{K_1}{I_1} \simeq \frac{K_2}{I_2}.
\end{equation}
Eq.(\ref{ratios}) comes from the analysis of the mass splittings for
the strange quark mass $m_{\rm s}=180$~MeV \cite{KimPraGo}. It
gets further support from the study of the model behavior in the
limit of the small soliton mass, which corresponds to the
non-relativistic quark model limit of the present model.
In this limit equality (\ref{ratios}) becomes exact.

Using (\ref{ratios}) we can write $c_{\overline{10}}$ and $c_{27}$
as follows:
\begin{equation}
m_{\rm s} c_{\overline{10}} = c,\;\;\;m_{\rm s} c_{27} = \frac 35 c,
\end{equation}
where $c$ is a parameter corresponding to the moment of inertia
$I_2$, which cannot be extracted from the mass splittings. Let us
remind that $I_2$ enters into the splittings between octet and {\em
exotic} baryonic states belonging to $\overline{10}$ or 27 \cite{exotic}.

The collective wave function can be explicitly written in terms of
the SU(3) Wigner $D^{({\cal R})}$ function:
\begin{equation}
\Psi^{(\cal R)}_B \;=\;  (-)^{S_3-1/2}\sqrt{\mbox{dim}(\cal R)}
\left[D^{(\cal R)}_{(YTT_3)(-1SS_3)} \right]^*.
\end{equation}
The flavor content of baryon magnetic moments is then expressed by
\begin{eqnarray}\label{uds}
\mu^{({\rm u})}_{B} &=& \mu^{(0)}_B + \frac{1}{2}\mu^{(3)}_B
                 +\frac{1}{2\sqrt{3}}\, \mu^{(8)}_B,     \nonumber \\
\mu^{({\rm d})}_{B} &=& \mu^{(0)}_B -  \frac{1}{2}\mu^{(3)}_B
                 +\frac{1}{2\sqrt{3}} \, \mu^{(8)}_B,     \nonumber \\
\mu^{({\rm s})}_{B} &=& \mu^{(0)}_B - \frac{1}{\sqrt{3}} \, \mu^{(8)}_B.
\label{Eq:flavor1}
\end{eqnarray}

Using Eqs.(\ref{Eq:mu},\ref{uds}),
we can easily express the magnetic moments of the octet:
\begin{equation}\label{mm8}
\mu_B \;=\;
\frac12 \left\langle B \left|  \left(\hat{\mu}^{(3)}
+ \frac{1}{\sqrt{3}} \hat{\mu}^{(8)} \right)\right|B\right\rangle
\;=\; \frac13 \left\langle B \left|  \left( 2 \hat{\mu}^{(\rm u)}
- \hat{\mu}^{(\rm d)} - \hat{\mu}^{(\rm s)}  \right)\right|B\right\rangle.
\end{equation}

With the accuracy ${\cal O}(m_{\rm s})$, we can express flavor
magnetic moments as follows:
\begin{equation}
\mu^{(\rm f)}_{B} \;=\; \mu^{(\rm f)}_B (m^0_{s})
+ \mu^{(\rm f)}_B (m^0_{s},\text{op})
+ \mu^{(\rm f)}_B (m^0_{s},\text{wf}),
\end{equation}
where f equals u, d or s.  $\mu^{(\rm f)}_B (m^0_{s}) $
denotes the SU(3) symmetric part of the magnetic moments, while the
symmetry breaking parts $\mu^{(\rm f)}_B (m^0_{s},\text{op})$ and
$\mu^{(\rm f)}_B (m^0_{s},\text{wf})$ correspond to the symmetry
breaking in the operator and in the baryon wave functions, respectively.

As in Ref.~\cite{KimPraGo}, we introduce a new set of parameters:
\begin{eqnarray}\label{vwxyzpq}
v&=& \frac{1}{60}\left(w_1 - \frac12 w_2\right),\;\;
w=\frac{1}{120} w_3, \nonumber \\
x&=& \frac{1}{540} m_{\rm s} w_4,\;\;\; y=\frac{1}{90} m_{\rm s} w_5,\;\;\;
z=\frac{1}{30} m_{\rm s} w_6, \nonumber \\
p&=&\frac16 c\left(w_1+w_2 + \frac12 w_3\right),\;\;\;
q=-\frac{1}{150} c\left(w_1+2w_2 - \frac32 w_3\right),
\end{eqnarray}
where $w_1=w_1^1+ m_{\rm s} w_1^2$.
We can now express Eq.(\ref{Eq:flavor1}) explicitly in terms
of the new variables of Eq.({\ref{vwxyzpq}):
\begin{eqnarray}
\mu^{({\rm u})}_{\rm p}= \mu^{({\rm d})}_{\rm n} & = &
 - 8\,v + 24\,w - 8\,x - 8\,y + 8\, q,
\nonumber \\
\mu^{({\rm u})}_{\rm n} = \mu^{({\rm d})}_{\rm p} & = &
~~ 6\,v + 22\,w + 14\,x + 2\,y + 2\,z + 2\,p + 4\, q,
\nonumber \\
\mu^{({\rm u})}_{\Lambda} = \mu^{({\rm d})}_{\Lambda} & = &
~~ 3\,v + 21\,w - 9\,x + 9\,y - 3\,z + 9\,q,
\nonumber \\
\mu^{({\rm u})}_{\Sigma^+} = \mu^{({\rm d})}_{\Sigma^-} & = &
- 8\,v + 24\,w - 4\,x - 10\,y + 4\,z + 4\,q,
\nonumber \\
\mu^{({\rm u})}_{\Sigma^0} = \mu^{({\rm d})}_{\Sigma^0} & = &
- 3\,v + 19\,w + 5\,x - 7\,y + 3\,z +  p + 4\,q,
\nonumber \\
\mu^{({\rm u})}_{\Sigma^-} = \mu^{({\rm d})}_{\Sigma^+} & =&
~~ 2\,v + 14\,w + 14\,x - 4\,y + 2\,z + 2\,p + 4\,q,
\nonumber \\
\mu^{({\rm u})}_{\Xi^0} = \mu^{({\rm d})}_{\Xi^-} & = &
~~ 6\,v + 22\,w - 4\,x + 11\,y - 5\,z + 4\,q,
\nonumber \\
\mu^{({\rm u})}_{\Xi^-} = \mu^{({\rm d})}_{\Xi^0} & = &
~~ 2\,v + 14\,w - 8\,x + 7\,y - 3\,z + 8\,q,
\nonumber \\
\mu^{({\rm s})}_{\rm p}=\mu^{({\rm s})}_{\rm n} &= &
~~ 2\,v + 14\,w - 6\,x - 3\,y + z -2\,p - 12\,q,
\nonumber \\
\mu^{({\rm s})}_{\Lambda} & = &
- 6\,v + 18\,w + 18\,x + 9\,y - 3\,z -18\,q,
\nonumber \\
\mu^{({\rm s})}_{\Sigma^{\pm} } =\mu^{({\rm s})}_{\Sigma^{0} } & = &
~~ 6\,v + 22\,w - 10\,x - 13\,y + 3\,z -2\,p - 8\,q,
\nonumber \\
\mu^{({\rm s})}_{\Xi^{0}} = \mu^{({\rm s})}_{\Xi^{-}} & = &
- 8\,v + 24\,w + 12\,x + 18\,y - 4\,z -12\,q.
\label{muds}
\end{eqnarray}
Equations (\ref{muds}) exhibit both isospin and "hypercharge"
symmetries of flavor magnetic moments.

In Ref.~\cite{KimPraGo} we have fitted parameters $v$, $w$, $x$, $y$,
$z$, $p$ and $q$ to the experimental data of the octet magnetic
moments.  It is of great convenience to fit first 2 parameters $v$ and
$w$ in such a way that they are independent of $m_{\rm
s}$\footnote{Strictly speaking  $v$ and $w$ can be extracted
independently of the remaining unknowns in Eq.(\ref{muds}), it is,
however, impossible to get rid of a weak $m_{\rm s}$ dependence in
$w_1=w_1^1+m_{\rm s}\, w_1^2$, which we neglect throughout this
paper.}:
\begin{equation}\label{vw}
\begin{array}{crl}
v=&-0.268 &= \left( 2
{\rm n}+3\Xi ^0+\Xi ^{-}-{\rm p}-2\Sigma ^{-}-3\Sigma ^{+}\right) /60, \\
w=&0.060 & =\left( 3
{\rm p}+4{\rm n}+\Xi ^0-3\Xi ^{-}-4\Sigma ^{-}-\Sigma ^{+}\right) /60.
\end{array}
\end{equation}
This procedure has the advantage that one can study the chiral limit of
the flavor magnetic moments, by tuning parameters $x,y,z,p$ and $q$ to
zero. As already mentioned, throughout our analysis we assume
that the mass of the strange quark is equal to 180~MeV.

Variable $z$ can also be extracted independently and the result reads:
\begin{equation}\label{z}
z=-0.080 =\frac16 ({\rm n}-\Sigma ^{-}+\Sigma ^{+}-\Xi ^0-{\rm p}+\Xi ^{-}).
\end{equation}
With these values for $v, w$ and $z$ we obtain 4 equations for the remaining
4 variables:
\begin{eqnarray} \label{xypq}
\left[ \begin{array}{rrrr}
        -8 & -5 & 0 & 8 \\
        14 &  5 & 2 & 4 \\
        -9 &  0 & 0 & 9 \\
        -4 & -1 & 0 & 4
        \end{array} \right] \left[ \begin{array}{c} x \\ y \\ p \\ q
                                   \end{array} \right]
 & = & \left[ \begin{array}{r} 0.326 \\ -0.343\\ 0.134 \\ 0.157
                                   \end{array} \right]; ~~~~~
       \left[ \begin{array}{c}
       {\rm p},~\Xi^- \\ {\rm n},~\Sigma^-\\ \Lambda \\ \Sigma^+, \Xi^0
                                   \end{array} \right].
\end{eqnarray}
Four equations (\ref{xypq}) are obtained by inserting (\ref{vw}) and
(\ref{z}) into the equations for the magnetic moments (\ref{muds}) and
(\ref{mm8}) (or equivalently into Eq.(50) of Ref.~\cite{KimPraGo}). They
are obtained from the equations for the magnetic moments of the
particles given in the last column of Eq.(\ref{xypq})\footnote{For example
the first equation in (\ref{xypq}) has been obtained from the equation
for proton and it is identical to the one obtained from the equation
for $\Xi^-$.}. By redefining $x=x^{\prime}-p/9$ and $q=q^{\prime}-p/9$ one
can get rid of the $p$ dependence. So in fact we are left with 4 equations
for 3 parameters: $x^{\prime}, ~y$ and $q^{\prime}$. In principle
one could expect to obtain 4 independent fits by removing one equation
from the set (\ref{xypq}) and then solving it for $x^{\prime}, ~y$ and
$q^{\prime}$.
However, by removing the equation for (${\rm n},~\Sigma^-$) we
get  a set of linearly-dependent equations, so we content
ourselves with only three fits\footnote{In Ref~\cite{KimPraGo} we have
presented results corresponding to fit (2).} obtaining:
\begin{eqnarray}\label{fits}
\begin{array}{lll}
{\rm ~~~~~~fit (1)} &{\rm ~~~~~~fit (2)} & {\rm ~~~~~~fit (3)} \\
{\rm p},~\Xi^-  {\rm ~~removed} &
~~\Lambda {\rm ~~removed} & ~~\Sigma^+, \Xi^0 {\rm ~~removed} \\
                  &                      &                   \\
x^{\prime}=~~0.005, &~~x^{\prime}=-0.026,    & ~~x^{\prime}=-0.012, \\
y=-0.097,          &~~y=-0.004,            & ~~y=-0.037,            \\
q^{\prime}=~~0.020, &~~q^{\prime}=~~0.012, & ~~q^{\prime}=~0.003.
\end{array}
\end{eqnarray}
The results, both for flavor and total magnetic moments are of course
independent of $p$. They are displayed in Table I. The fits (1), (2) and (3)
of  Eq.(\ref{fits}) differ by $\chi^2$ which equals 2.78, 1.66 and 0.058
respectively. Fit (3) is by far the best;
it misses $\Sigma^+$ by 2\% and $\Xi^0$ by 7\%. Guided
by this accuracy we have minimized $\chi^2$ assuming that the
average error is 15\%. One should interpret this error as the theoretical
error corresponding to the truncation of the perturbation series in 
$m_{\rm s}$. The result reads:
\begin{eqnarray}\label{fit4}
\begin{array}{ccc}
 & {\rm fit(4)} &   \\
 &              &    \\
v=-0.266 \pm 0.030, & w= 0.064 \pm 0.036, & x^{\prime} = -0.008 \pm 0.017, \\
y=-0.050 \pm 0.042, & z=~0.093 \pm 0.190, & q^{\prime} =~~0.006 \pm 0.020,
\end{array}
\end{eqnarray}
corresponding to $\chi^2=0.017$. Errors have been calculated by allowing
$\chi^2$ to change by 1.
Results of fit (4) are given in Table I. In Table I we also
present the model calculation ($\chi$QSM) for the constituent mass
$M=420$~MeV and $m_{\rm s}=180$~MeV and (where available) the results
of the model independent fit 1 of Ref.~\cite{HongParkMin} (HPM).

It should be mentioned that the non-relativistic
limit of the chiral quark--soliton model
(which, to some extent, can be used as a guiding line) predicts
 $w/v= - 1/7$,
$3 y/v=z/v=8/21\, m_s I_2$, and
$q^{\prime} =x^{\prime} = 0$
which is in a qualitative agreement
with our fit eq.~(\ref{fit4}). With values of the constants in
non-relativistic limit we get:
\begin{equation}
\frac{\mu_p}{\mu_n}=-\frac 32 (1-\frac{8}{135} m_s I_2),\qquad
\mu_s=0\, ,
\end{equation}
these are the well known results of the $SU(6)$ non-relativistic quark model
with particular mass correction to them (note that mass corrections
for $\mu_s$ are exactly zero in the non-relativistic limit).

Before we comment on the results presented in Table I let us discuss the
sensitivity of our fits to the symmetry breaking. To do this we
reconsider fits (1), (2) and (3) and restore the linear $m_{\rm s}$
dependence of the variables $x^{\prime},~y,~z,$ and $q^{\prime}$.
At this point we can profit from our fitting procedure in which chiral
limit parameters $v$ and $w$ were fitted independently of $m_{\rm s}$.
The results for fit (3) are plotted in Fig.1.  We can see from Fig.1
that $\mu^{({\rm s})}_B$ is  rather stable, as far as $m_{\rm s}$
dependence is concerned. ``Error bars'' in Fig.1.
correspond to the range of values for $\mu^{({\rm s})}_B (m_{\rm
s}=180~{\rm MeV})$ for all fits (with fit (4) included). However, as can
be seen from Fig.1, the nucleon line is almost flat:
$\mu^{({\rm s})}_{\rm N}$ is quite insensitive to the symmetry breaking.

Of course the value $\mu_{\rm N}^{({\rm s})}=0.41\, \mu_{\rm N}$
of fit (4) has its own error, which can be estimated by varying parameters
$v$, $w$, $x^{\prime}$, $y$, $z$ and $q^{\prime}$ within the
error bars given in Eq.(\ref{fit4}). The average error obtained
by this procedure equals to $\pm 0.18$ assuming 15\% accuracy in the
$\chi^2$ fit (or $\pm 0.12$ for 10 \% accuracy).

Despite the uncertainties for the strange particles we conclude
that the data suggest rather large strangeness contribution to the
magnetic moment of the nucleon. On the contrary, the self-consistent
$\chi$QSM gives for $\mu^{(\rm s)}_{\rm p,n}$ a small value, which
is close to zero, but positive~\footnote{Note that Ref.~\cite{Ki} includes
a numerical error in the calculation of the strange magnetic
moment.}. This is due to the fact that parameter $w_3$ (or equivalently
$w$) is much smaller than the one obtained by our fitting procedure.
As a result the ``starting point'' values for magnetic moments (at
$m_{\rm s}$=0) differ from the ones obtained by all our ``model
independent'' fits with the help of Eq.(\ref{vw}). In the self-consistent
$\chi$QSM linear $m_{\rm s}$ dependence is not strong enough to change the
chiral limit results enough to reproduce the experimental data for the
magnetic moments. Also the strange magnetic moment of the nucleon stays close
to the SU(3) symmetry value which is almost zero (but negative).

\vspace{0.8cm}
\centerline{\epsfysize=3.3in\epsffile{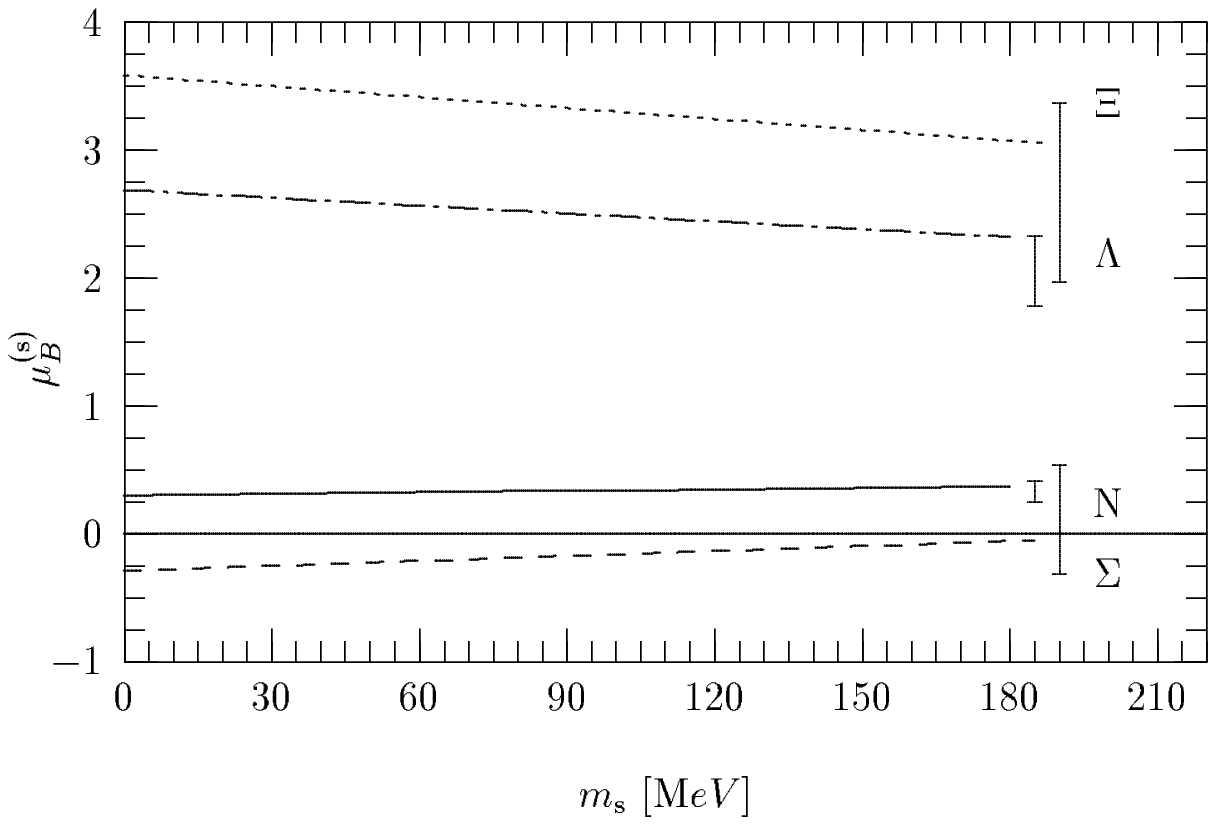}}\vskip4pt
\begin{center}
{\footnotesize {\bf Fig. 1}:
~$m_{\rm s}$ dependence of the strange magnetic moments.}
\end{center}

Our present ``model-independent''
results for $\mu^{(\rm u)}_{\rm p,n}$ and $\mu^{(\rm s)}_{\rm p,n}$
agree quite well with the ones of Ref.~\cite{HongParkMin}.
Numerically the contribution of the terms proportional to the
difference $w_5-w_6$ is quite small; in the worse case of fit (1) it
does not exceed 8 \% of the proton magnetic moment.

In summary, we have investigated the flavor structure of
the octet magnetic moments within the framework of the
chiral quark-soliton model, employing a ``model-independent''
approach.  The strange magnetic moment of the nucleon turns out
to be positive.  The effect of the SU(3) symmetry breaking
is almost negligible in the case of the nucleon.
The ``model-independent'' fit
($\mu^{({\rm s})}_{\rm N}\simeq (0.25 - 0.37)\, \mu_{\rm N}$)
is stable and yields quite larger value than that of  the
self-consistent model calculation
( $ \mu^{({\rm s})}_{\rm N} = 0.03  \mu_{\rm N}  $).
On the other hand the best fit obtained by minimizing $\chi^2$
assuming 15\% theoretical accuracy yields
$\mu^{({\rm s})}_{\rm N} = (0.41 \pm 0.18 )\; \mu_{\rm N}   $.

This work has partly been supported by the BMBF, the DFG
and the COSY--Project (J\" ulich).  M.V.P. and M.P. have been
supported by Alexander von Humboldt Stiftung.

\newpage

\thispagestyle{empty}

\vspace{-2cm}


\noindent
Table 1: Magnetic moments and their flavor decomposition for 4 fits
described in the text, for model calculations ($\chi$QSM) and for fit 1
of Ref.\cite{HongParkMin}(HPM).
\vspace{0.2cm}

\begin{tabular}{|c|c|r|r|r|c|}
\hline
~particle~ & fit/model & ~~~~$\mu^{(\rm u)}$~~~~ &
                       ~~~~$\mu^{(\rm s)}$~~~~ &
                       ~~~~$\mu_{\rm th} $~~~~ &
                       ~~~~$\mu_{\rm exp}$~~~~ \\
\hline
     & (1) &  4.48 & 0.25 & 3.05 &        \\
     & (2) &  3.92 & 0.25 & inp. &        \\
 p   & (3) &  4.00 & 0.37 & inp. & 2.79   \\
     & (4) &  4.18 & 0.41 & 2.84 &         \\
     & $\chi$QSM &  3.02 & 0.03 & 2.39 &        \\
     & HPM  &  4.08 & 0.33 & 2.80 &        \\
\hline
     & (1) & -0.50 &      & inp. &        \\
     & (2) & -0.78 &      & inp. &        \\
 n   & (3) & -0.68 & as p & inp. & -1.91  \\
     & (4) & -0.55 &      & -1.89&        \\
     & $\chi$QSM & -1.15 &      & -1.78&        \\
     & HPM  & -0.63 &      & -1.91&        \\
\hline
           & (1) & -0.05 & 1.78 & inp. &        \\
           & (2) &  1.00 & 2.20 &-0.40 &        \\
$\Lambda$  & (3) &  0.50 & 2.32 & inp. & -0.61  \\
           & (4) &  0.51 & 2.33 & -0.61&         \\
           & $\chi$QSM & -0.29 & 1.94 &-0.74 &        \\
\hline
           & (1) &  4.30 & 0.54 & inp. &        \\
           & (2) &  3.46 &-0.31 & inp. &        \\
$\Sigma^+$ & (3) &  3.70 & 0.05 & 2.40 &  2.46  \\
           & (4) &  3.85 & 0.23 & 2.34 &         \\
           & $\chi$QSM &  3.18 &-0.75 & 2.46 &        \\
\hline
           & (1) &  2.49 &              & 0.65 &        \\
           & (2) &  1.65 &              & 0.65 &        \\
$\Sigma^0$ & (3) &  1.92 & as $\Sigma^+$& 0.62 & no data\\
           & (4) &  2.07 &              & 0.59 &         \\
           & $\chi$QSM &  1.46 &              & 0.74 &        \\
\hline
           & (1) &  0.68 &              & inp. &        \\
           & (2) & -0.16 &              & inp. &        \\
$\Sigma^-$ & (3) &  0.14 & as $\Sigma^+$& inp. & -1.16  \\
           & (4) &  0.29 &              &-1.17 &         \\
           & $\chi$QSM & -0.26 &              &-0.98 &        \\
\hline
           & (1) & -0.90 & 1.97 & inp. &        \\
           & (2) &  0.22 & 3.37 & inp. &        \\
$\Xi^0  $  & (3) & -0.24 & 3.05 &-1.31 & -1.25  \\
           & (4) & -0.21 & 2.98 &-1.27 &         \\
           & $\chi$QSM & -1.24 & 2.69 &-1.65 &        \\
\hline
           & (1) & -0.02 &           &-0.37 &        \\
           & (2) &  0.82 &           & inp. &        \\
$\Xi^-  $  & (3) &  0.40 &as $\Xi^0$ & inp. & -0.65  \\
           & (4) &  0.41 &           &-0.65 &         \\
           & $\chi$QSM & -0.22 &           &-0.63 &        \\
\hline
\end{tabular}



\begin{thebibliography}{99}
\bibitem{GasserLeutwylerSainio} J. Gasser, H. Leutwyler and
M.E. Sainio, {\em Phys. Lett.} {\bf B253} (1991) 252.
\bibitem{ashman} J. Ashman {\em et al}., {\em Nucl. Phys.}
{\bf B328} (1989) 1.
\bibitem{km} D.B. Kaplan and A. Manohar, {\em Nucl. Phys.}
{\bf B310} (1988) 527.
\bibitem{Ja} R.L. Jaffe, {\em Phys. Lett.} {\bf 229B}
(1989) 275.
\bibitem{Pa} N.W. Park, J. Schechter and H. Weigel,
{\em Phys. Rev.} {\bf D43} (1991) 869.
\bibitem{pw} N.W. Park and H. Weigel, {\em Nucl. Phys.}
{\bf A541} (1992) 453.
\bibitem{Ko} W. Koepf, E.M. Henley and S.J. Pollock,
{\em Phys. Lett.} {\bf 288B} (1992) 11.
\bibitem{Co} T.D. Cohen, H.Forkel and M. Nielsen, {\em Phys. Lett.}
{\bf B316} (1993) 1.
\bibitem{Ho} S.T. Hong and B.-Y. Park, {\em Nucl. Phys.} {\bf A561},
525 (1993).
\bibitem{mu} M.J. Musolf and M. Burkardt, {\em Z. Phys.}
{\bf C61} (1994) 433.
\bibitem{fo}H. Forkel, M. Nielsen, X. Jin and T.D. Cohen,
{\em Phys. Rev.} {\bf C50} (1994) 3108.
\bibitem{Le} D.B. Leinweber, {\em Phys.Rev.} {\bf D53} (1996) 5115.
\bibitem{Ha} H.W. Hammer, Ulf-G. Mei\ss ner and D. Drechsel,
{\em Phys.Lett.} {\bf B367} (1996) 323.
\bibitem{Mus} M.J. Musolf and H. Ito, DOE-ER-40561-245
[{\tt nucl-th/9607021}], (1996).
\bibitem{Ki} H.-C. Kim, T. Watabe and K. Goeke, {\em Nucl. Phys}
{\bf A616} (1997) 606.
\bibitem{Me} Ulf-G. Meissner, V. Mull, J. Speth and
J.W. van Orden, {\em Phys. Lett.} {\bf B408} (1997) 381.
\bibitem{HongParkMin} S.-T. Hong, B.-Y. Park and D.-P. Min,
SNUTP-97-032 [{\tt nucl-th/9706008}], (1997).
\bibitem{SAMPLE} M. Mueller {\em et al.}, {\em Phys. Rev. Lett.} {\bf 78}
(1997) 3824.
\bibitem{CEBAF1} CEBAF proposal $\#$ PR-91-017 (1991),
D.H. Beck, spokesperson.
\bibitem{CEBAF2} CEBAF proposal $\#$ PR-91-004 (1991),
E.J. Beise, spokesperson.
\bibitem{CEBAF3} CEBAF proposal $\#$ PR-91-010 (1991), M. Finn and
P.A. Souder, spokespersons.
\bibitem{Mainz} Mainz proposal A4/1-93 (1993), D. von Harrach, spokesperson.
\bibitem{MIT1} MIT-Bates proposal $\#$ 89-06 (1989), R.D. McKeown and
D.H. Beck, spokespersons.
\bibitem{MIT2} MIT-Bates proposal $\#$ 94-11 (1994), M. Pitt and E.J. Beise,
spokespersons.
\bibitem{KPBG}H.--C. Kim, M.V. Polyakov, A. Blotz and K. Goeke,
{\em Nucl. Phys.} {\bf A598}, 379 (1996).
\bibitem{WaKaya} M. Wakamatsu and N. Kaya, {\em Progr. of Theor. Phys. }%
{\bf 95}, 767 (1996).
\bibitem{KimPraGo} H.-C. Kim, M. Prasza\l owicz and K. Goeke,
RUB-TPII-19/96,[{\tt hep-ph/9706531}], (1997),
to be published in {\em Phys. Rev.} {\bf D}, (1998).
\bibitem{review} Chr. V. Christov, A. Blotz, H.-C. Kim,
P. Pobylitsa, T. Watabe, Th. Meissner, E. Ruiz Arriola and K. Goeke,
{\em Prog. Nucl. Part. Phys.} {\bf 37} (1996) 91.
\bibitem{Blotzetal}A. Blotz, D. Diakonov, K. Goeke, N.W. Park,
V. Petrov and P.V. Pobylitsa, {\em Nucl. Phys.} {\bf A555} (1993) 765.
\bibitem{exotic} D. Diakonov, V. Petrov and M. Polyakov,
{\em Z.Phys.} {\bf A359} (1997) 305.
\end{thebibliography}
\end{document}